\documentclass[final,5p,times]{elsarticle}

\usepackage{amsmath}
\usepackage{amssymb}
\usepackage{graphicx}
\usepackage{float}
\usepackage{placeins}
\usepackage{booktabs}
\usepackage{siunitx}
\usepackage{xspace}
\graphicspath{{./}}

\newcommand{\ddme}{DD-ME2\xspace}
\newcommand{\ddpc}{DD-PC1\xspace}
\newcommand{\fv}{f_V}
\newcommand{\xt}{x_t}
\newcommand{\wt}{w_t}

\begin{document}

\begin{frontmatter}

\title{Finite-nucleus-protected high-density extension of covariant density functionals constrained by multimessenger data}

\author[addr1,addr2]{Wen-Jie Xie\corref{cor1}}
\ead{xiewenjie@ycu.edu.cn}
\author[addr1,addr2]{Jun-Hua Guo}
\cortext[cor1]{Corresponding author}
\address[addr1]{Department of Physics, Yuncheng University, Yuncheng 044000, China}
\address[addr2]{Shanxi Province Intelligent Optoelectronic Sensing Application Technology Innovation Center, Yuncheng University, Yuncheng 044000, China}

\begin{abstract}
We construct a finite-nucleus-protected high-density extension of covariant
density functionals by modifying only the isoscalar-vector channel outside
the finite-nucleus calibration domain.  The extension introduces three
parameters controlling the strength, onset, and width of the high-density
deformation, while the scalar and isovector channels are kept unchanged.  A
Bayesian analysis using heavy-ion flow constraints, massive-pulsar
information, NICER mass-radius measurements, and the GW170817 tidal
constraint shows that the original \ddme interaction is strongly disfavored
relative to its protected high-density extension, with
\(\ln K=\ln(Z_{\rm ext}/Z_{\rm base})=26.67\), where \(Z\) denotes the
Bayesian evidence, after imposing a causal/stability filter on the reshaped
EOS.  In contrast, \ddpc serves as a reference functional for which the same
extension is not required by the present data, giving \(\ln K=-0.44\).  The
result supports the interpretation that the proposed extension is not an
unconstrained phenomenological patch: Bayesian evidence selects it only when
demanded by the combined high-density data, while finite-nucleus observables
remain unchanged within numerical precision.
\end{abstract}


\end{frontmatter}

\section{Introduction}

The equation of state (EOS) of cold dense matter is now constrained by a
combination of laboratory, electromagnetic, and gravitational-wave data.
Classic constraints from collective flow in heavy-ion collisions first
established pressure bands for symmetric matter at several times nuclear
saturation density~\cite{Danielewicz2002}.  Recent heavy-ion
programs and transport-model analyses have renewed this effort in the
multimessenger era~\cite{Sorensen2024PPNP,OmanaKuttan2023PRL,Tsang2024NatAstron},
while astrophysical observations have added complementary information from
massive pulsars~\cite{Antoniadis2013,Cromartie2020,Romani2022},
GW170817~\cite{Abbott2017PRL,Abbott2018PRL}, and NICER
mass-radius measurements~\cite{Riley2021,Miller2021}.
The newest NICER analyses of PSR J0437--4715 and related EOS updates further
tighten this data set~\cite{Choudhury2024,Rutherford2024,Brandes2025}.
The same thermodynamic information can also be expressed through the trace
anomaly \(\Delta=1/3-P/\varepsilon\), which has recently been used to connect
sound-speed structure in neutron-star matter with the approach to conformal
behavior~\cite{Fujimoto2022PRL}.  A complementary extraction from
intermediate-energy collective flow further suggests that HIC and
neutron-star data constrain mutually consistent macroscopic stiffness
measures~\cite{Li2026TraceAnomalyFlow}.

These developments have made Bayesian EOS inference a central tool for dense
matter physics~\cite{Raaijmakers2021,Legred2021,Huth2022,
Brandes2023,Essick2023,Jiang2023,Pang2024,Rutherford2024}.  At the same time,
microscopic calculations based on chiral effective field theory and
perturbative QCD increasingly constrain the low- and asymptotically
high-density limits~\cite{Drischler2021,Drischler2022,Keller2023PRL,
Komoltsev2022,Gorda2023JHEP,Gorda2023ApJ}.  The intermediate-density region
relevant for massive neutron stars remains less certain, and recent analyses
have discussed possible high-density softening, trace-anomaly behavior,
quark-matter signatures, and speed-of-sound structure~\cite{Annala2020,
Altiparmak2022,Annala2023,Brandes2023,Marczenko2023,Takatsy2023,
Mroczek2024,Lim2024,Kaiser2024}.

Covariant density functionals (CDFs) are a natural framework for connecting
finite nuclei and dense matter~\cite{Serot1986,Vretenar2005,
Niksic2011}.  Density-dependent meson-exchange and point-coupling
functionals such as \ddme and \ddpc have been calibrated with high accuracy
to finite nuclei~\cite{Typel1999,Lalazissis2005,Niksic2008}, and related
Bayesian studies have examined how relativistic mean-field or CDF parameters
are constrained by nuclear and astrophysical data~\cite{Salinas2023}.
However, finite nuclei probe densities near
and below saturation, whereas HIC flow and neutron-star cores probe several
times saturation density.  A direct global refit can therefore entangle the
well-calibrated finite-nucleus sector with the less constrained high-density
sector.

Here we adopt a different strategy.  We introduce a finite-nucleus-protected
high-density extension of the functional and use Bayesian evidence to decide
whether this extension is required by the data.  The main case is \ddme,
which provides an accurate finite-nucleus functional but produces a relatively
stiff high-density EOS.  The reference case is \ddpc, which is treated with
the same extension and likelihood but acts as a negative control: if the
method merely rewarded additional flexibility, the extension would be favored
for both functionals.  Instead, the Bayesian evidence selects the protected
high-density completion for \ddme but not for \ddpc, thereby using the same
framework to diagnose whether a given functional requires additional
high-density flexibility.  The goal is therefore not to present a completed
global refit of a universal CDF, but to isolate the high-density sector in a
controlled setting before finite-nucleus observables are incorporated
directly into a future global likelihood.

\section{Framework and Bayesian analysis}

The extension is deliberately minimal.  We interpret it as a
finite-nucleus-protected high-density counterterm in the isoscalar-vector
channel, or equivalently as an in-medium renormalization of the vector
self-energy.  This channel provides the most direct lever arm on the
short-range repulsion and hence on the stiffness of dense matter.  The scalar
and isovector channels are kept fixed,
\begin{equation}
  f_S=f_{TS}=f_{TV}=0,
\end{equation}
so that the low-density balance responsible for finite nuclei is not refitted.
This separation follows the structure of relativistic mean-field and density
functional descriptions, where the scalar-vector balance controls saturation
and finite nuclei, while the high-density vector sector determines the
repulsive contribution to the pressure~\cite{Serot1986,Vretenar2005,
Niksic2011}.  The baseline interactions are the density-dependent
meson-exchange functional \ddme~\cite{Lalazissis2005} and the
density-dependent point-coupling functional \ddpc~\cite{Niksic2008}.
The choice \(f_{TV}=0\) is part of the model definition rather than an
assertion that the high-density symmetry energy is unimportant.  The HIC flow
constraint used here acts directly on symmetric nuclear matter, where the
isovector-vector contribution vanishes.  Opening a high-density isovector
counterterm would therefore mix the present diagnosis of the isoscalar
pressure with a poorly constrained symmetry-energy inference.  We leave that
larger parameter space to a future global analysis including finite-nucleus
isovector data and asymmetric-matter HIC constraints.
In terms of the baryon density \(n\) and \(x=n/n_0\), the vector-channel
deformation is implemented as
\begin{equation}
  \begin{aligned}
  \Gamma_V(n;\boldsymbol{\theta}) &=
  \Gamma_V^{(0)}(n)\,
  \left[1+\fv\,S(x;\xt,\wt)\right],\\
  \boldsymbol{\theta} &= (\fv,\xt,\wt).
  \end{aligned}
\end{equation}
where \(\Gamma_V^{(0)}\) denotes the original \ddme or \ddpc vector coupling.
The modification is made at the level of the density-dependent coupling; the
corresponding vector self-energy and rearrangement self-energy are then
recomputed consistently in the EOS.
For the favored \ddme solutions the posterior selects negative \(\fv\).  This
should not be read as a literal removal of the underlying vector interaction;
rather, within a density-dependent CDF it is an effective reduction of the
net high-density isoscalar-vector repulsion after unresolved many-body
correlations, medium screening, exchange contributions, or emergent
non-nucleonic degrees of freedom have been integrated out.
The dimensionless switching function used in the numerical implementation is
\begin{equation}
  S(x;\xt,\wt)=H(x)\,
  \frac{1+\tanh[(x-\xt)/\wt]}{2},
\end{equation}
with the finite-nucleus protection factor
\begin{equation}
  H(x)=
  \begin{cases}
    0, & x\le x_{\rm cut},\\
    3t^2-2t^3, & x_{\rm cut}<x<x_{\rm on},\\
    1, & x\ge x_{\rm on},
  \end{cases}
  \quad
  t=\dfrac{x-x_{\rm cut}}{x_{\rm on}-x_{\rm cut}} .
\end{equation}
We use \(x_{\rm cut}=1.35\) and \(x_{\rm on}=2.00\).  These two scales are
treated as protection scales, not as inference parameters in the present
Letter.  The value \(x_{\rm cut}=1.35\) lies above the density region
effectively sampled by finite nuclear ground states and provides a safety
margin for the finite-nucleus calibration, while \(x_{\rm on}=2.00\) marks
the beginning of the density interval where HIC flow constraints become
directly restrictive.  Thus the deformation is strictly switched off in the
finite-nucleus calibration domain, is smoothly turned on above
\(x_{\rm cut}\), and approaches an asymptotic fractional change \(\fv\) in
the vector channel at high density.  The onset parameter \(\xt\) and width
\(\wt\) are varied in the Bayesian analysis.
Varying \(x_{\rm cut}\), \(x_{\rm on}\), or the functional form of the switch
would define a larger model space.  In the main analysis we keep these
protection scales fixed in order to test the minimal three-parameter
extension, and below we quantify the sensitivity of the \ddme conclusion to
moderate changes of \(x_{\rm cut}\).

The prior range is intentionally broad,
\begin{equation}
  \fv\in[-0.20,0.20],\qquad
  \xt\in[1.35,5.00],\qquad
  \wt\in[0.40,3.00].
\end{equation}
Here \(\fv\) controls the asymptotic strength of the high-density deformation,
while \(\xt\) and \(\wt\) determine its onset and width.  The no-reshape
baselines are included as zero-parameter models and are evaluated with the
same likelihood as the reshaped models.

The construction is phenomenological in the sense that it represents
unresolved high-density many-body physics.  It is, however, not an
unconstrained patch.  The deformation is restricted to the high-density
vector sector, leaves the finite-nucleus calibration sector unchanged, and
is penalized by Bayesian evidence through the additional prior volume.  The
density-dependent rearrangement contribution associated with the modified
vector coupling is included consistently, as required for thermodynamic
consistency in density-dependent relativistic mean-field models
~\cite{Typel1999}.  As a direct audit, finite-nucleus calculations were
repeated for the tested high-density cases.  Binding energies, charge radii,
and neutron skins remain unchanged with respect to the original interactions
within numerical precision, confirming that the extension affects the
high-density EOS sector rather than the finite-nucleus fit.

\begin{table}[!t]
  \centering
  \caption{Finite-nucleus protection audit for the tested high-density
  deformations.  The entries give the largest absolute change over the
  finite-nucleus test set, \(^{48}\)Ca, \(^{68}\)Ni, \(^{90}\)Zr,
  \(^{132}\)Sn, and \(^{208}\)Pb, relative to the corresponding original
  interaction.  The differences are zero in the stored output files; we quote
  them as numerical upper limits rather than as exact physical zeros.}
  \label{tab:finite-protection}
  \begin{tabular}{lccc}
    \toprule
    family & \(\max|\Delta E/A|\) & \(\max|\Delta r_{\rm ch}|\) &
    \(\max|\Delta r_{\rm skin}|\) \\
     & (MeV) & (fm) & (fm) \\
    \midrule
    \ddme & \(<10^{-6}\) & \(<10^{-6}\) & \(<10^{-6}\) \\
    \ddpc & \(<10^{-6}\) & \(<10^{-6}\) & \(<10^{-6}\) \\
    \bottomrule
  \end{tabular}
\end{table}

\paragraph{Bayesian model comparison}

The likelihood used in the evidence calculation follows the implementation in
the common analysis scripts.  It combines the HIC flow constraint on the
pressure of symmetric nuclear matter~\cite{Danielewicz2002} with
neutron-star observables: a massive-pulsar lower bound on the maximum
mass~\cite{Antoniadis2013,Cromartie2020,Romani2022}, NICER mass-radius
likelihoods for PSR J0030+0451, PSR J0437--4715, and PSR
J0740+6620~\cite{Riley2019,Miller2019,Riley2021,Miller2021,
Choudhury2024,Rutherford2024}, and the GW170817 tidal
likelihood~\cite{Abbott2017PRL,Abbott2018PRL,
Abbott2019PRX}.  We do not impose a chiral effective-field theory likelihood
in the evidence calculation.  The cited chiral-EFT results are used only to
place the low-density regime in context and to motivate why the present
diagnostic focuses on densities above the finite-nucleus calibration domain
~\cite{Drischler2022,Keller2023PRL}.  A quantitative chiral-EFT likelihood
would require selecting a specific many-body calculation, regulator setup,
and covariance prescription; we leave that choice to a future global
analysis in which low-density microscopic information and finite-nucleus data
are treated in a common likelihood.  The total log likelihood is
\begin{equation}
  \begin{aligned}
  \ln {\cal L}(\boldsymbol{\theta}) &=
  \ln {\cal L}_{\rm HIC}
  +\ln {\cal L}_{M_{\max}}\\
  &\quad
  +\ln {\cal L}_{\rm NICER}
  +\ln {\cal L}_{\rm GW170817}.
  \end{aligned}
\end{equation}
Each term is evaluated in the same way as in the analysis scripts.  For the
HIC flow band, the model pressure \(P_{\rm SNM}(u_j)\) is sampled at 25
equally spaced points in the common domain of the digitized lower and upper
flow boundaries, restricted to \(1.3\le u=n/n_0\le4.5\).  In the present
files this gives \(u_j\in[1.3,4.5]\).  If
\(P_j^-\) and \(P_j^+\) are the lower and upper flow boundaries, we use a
one-sided Gaussian penalty,
\begin{equation}
  \ln {\cal L}_{\rm HIC}
  =-\frac{1}{2}\sum_j
  \left[
  \frac{\Delta P_j}{\sigma_j}
  \right]^2,
  \quad
  \sigma_j=\frac{P_j^+-P_j^-}{2},
\end{equation}
where
\begin{equation}
  \Delta P_j =
  \begin{cases}
    P_j^- - P_{\rm SNM}(u_j), & P_{\rm SNM}(u_j)<P_j^-,\\
    P_{\rm SNM}(u_j)-P_j^+, & P_{\rm SNM}(u_j)>P_j^+,\\
    0, & P_j^-\le P_{\rm SNM}(u_j)\le P_j^+ .
  \end{cases}
\end{equation}
This implementation treats the digitized flow band as a phenomenological
constraint and does not include a covariance matrix for point-to-point
correlations or transport-model systematic uncertainties.
The maximum-mass term is a lower-bound likelihood with
\(M_{\rm ref}=2.08\,M_\odot\) and \(\sigma_M=0.07\,M_\odot\),
\begin{equation}
  \ln {\cal L}_{M_{\max}}
  =-\frac{1}{2}
  \left[
  \frac{\max(0,M_{\rm ref}-M_{\max})}{\sigma_M}
  \right]^2 .
\end{equation}
For NICER, each source \(s\) is represented by a two-dimensional KDE
\(\widehat p_s(R,M)\) constructed from the corresponding mass-radius samples.
The model likelihood is the KDE density averaged along the stable
mass-radius curve.  This is the one-dimensional marginalization appropriate
when an EOS predicts a single stable \(R(M)\) branch through a two-dimensional
source posterior,
\begin{equation}
  {\cal L}_{\rm NICER}
  =\prod_s
  \frac{1}{M_{s,+}-M_{s,-}}
  \int_{M_{s,-}}^{M_{s,+}}
  \widehat p_s\!\left[R_{\boldsymbol{\theta}}(M),M\right]\,dM ,
\end{equation}
where \(M_{s,-}\) and \(M_{s,+}\) are the 0.5 and 99.5 percentile masses of
the source samples.  The product runs over PSR J0030+0451, PSR J0437--4715,
and PSR J0740+6620.

The GW170817 term uses the processed low-spin posterior in
\((q,\Lambda_1,\Lambda_2)\).  At fixed chirp mass
\({\cal M}=1.186\,M_\odot\), the component masses are
\begin{equation}
  m_1(q)={\cal M}(1+q)^{1/5}q^{-3/5},\qquad
  m_2(q)={\cal M}(1+q)^{1/5}q^{2/5}.
\end{equation}
With \(\widehat p_{\rm GW}\) denoting the KDE posterior density and
\(\pi_q(q|{\cal M})\propto (1+q)^{2/5}q^{-6/5}\) the fixed-\({\cal M}\)
mass-ratio prior used in the correction, the likelihood is evaluated as
\begin{equation}
  {\cal L}_{\rm GW170817}
  =
  \int_{q_{\min}}^{q_{\max}}
  \frac{
  \widehat p_{\rm GW}\!\left[
  q,\Lambda_{\boldsymbol{\theta}}(m_1(q)),
  \Lambda_{\boldsymbol{\theta}}(m_2(q))
  \right]}
  {\pi_q(q|{\cal M})}\,dq ,
\end{equation}
with \(q_{\min}=0.60\), \(q_{\max}=0.995\), and the numerical KDE evaluated
in the transformed variables
\((\log[q/(1-q)],\log\Lambda_1,\log\Lambda_2)\), including the corresponding
Jacobian.  Configurations outside the tidal-deformability support
\(0\le\Lambda_{1,2}\le 2\times10^4\) are assigned negligible likelihood.
The same HIC+neutron-star likelihood definition is used for all four models:
\begin{equation}
  \begin{aligned}
  &\mathrm{DD\mbox{-}ME2},\quad
  \mathrm{DD\mbox{-}ME2}+\mathrm{reshape},\\
  &\mathrm{DD\mbox{-}PC1},\quad
  \mathrm{DD\mbox{-}PC1}+\mathrm{reshape}.
  \end{aligned}
\end{equation}
For the reshaped models, posterior sampling and evidence evaluation are
performed in the three-dimensional parameter space \((\fv,\xt,\wt)\).  The
Bayesian evidence is
\begin{equation}
  Z_{\cal M} =
  \int d\boldsymbol{\theta}\,
  {\cal L}(D|\boldsymbol{\theta},{\cal M})
  \pi(\boldsymbol{\theta}|{\cal M}),
\end{equation}
where \({\cal M}\) denotes the model and \(D\) the combined HIC+neutron-star
data set.  For the
no-reshape models the parameter space is zero-dimensional, so \(Z_{\cal M}\)
reduces to the likelihood of the corresponding baseline.  The Bayes factor
between the reshaped and original versions of a given functional is therefore
\begin{equation}
  \ln K_{\rm reshape} =
  \ln Z_{\rm reshape}-\ln Z_{\rm original}.
\end{equation}
This construction makes the Occam penalty explicit: an extension is favored
only if its improved likelihood compensates for the additional prior volume.
This model-selection logic follows recent multimessenger EOS analyses, where
additional flexibility is meaningful only when supported by the data
~\cite{Raaijmakers2021,Legred2021,Brandes2023,Pang2024}.

The evidence values quoted below are obtained from MultiNest/PyMultiNest
runs in the three reshaping parameters for each extended model, using the
uniform prior ranges specified above.  We repeated the nested sampling with
independent random seeds and \(n_{\rm live}=2000\) live points; the reported
\(\ln Z\) values are the run averages, while the run-to-run scatter and the
mean internal MultiNest error are reported in Table~\ref{tab:evidence}.  The
same likelihood code is also evaluated on deterministic grids to check prior
and protection-scale sensitivities.  These grid calculations are not used as
the final evidence values, but they provide reproducibility checks on the
sign and scale of the Bayes factors and on the effect of imposing the
causal/stability filter.

\section{Results and discussion}

\begin{figure*}[!t]
  \centering
  \includegraphics[width=0.92\linewidth]{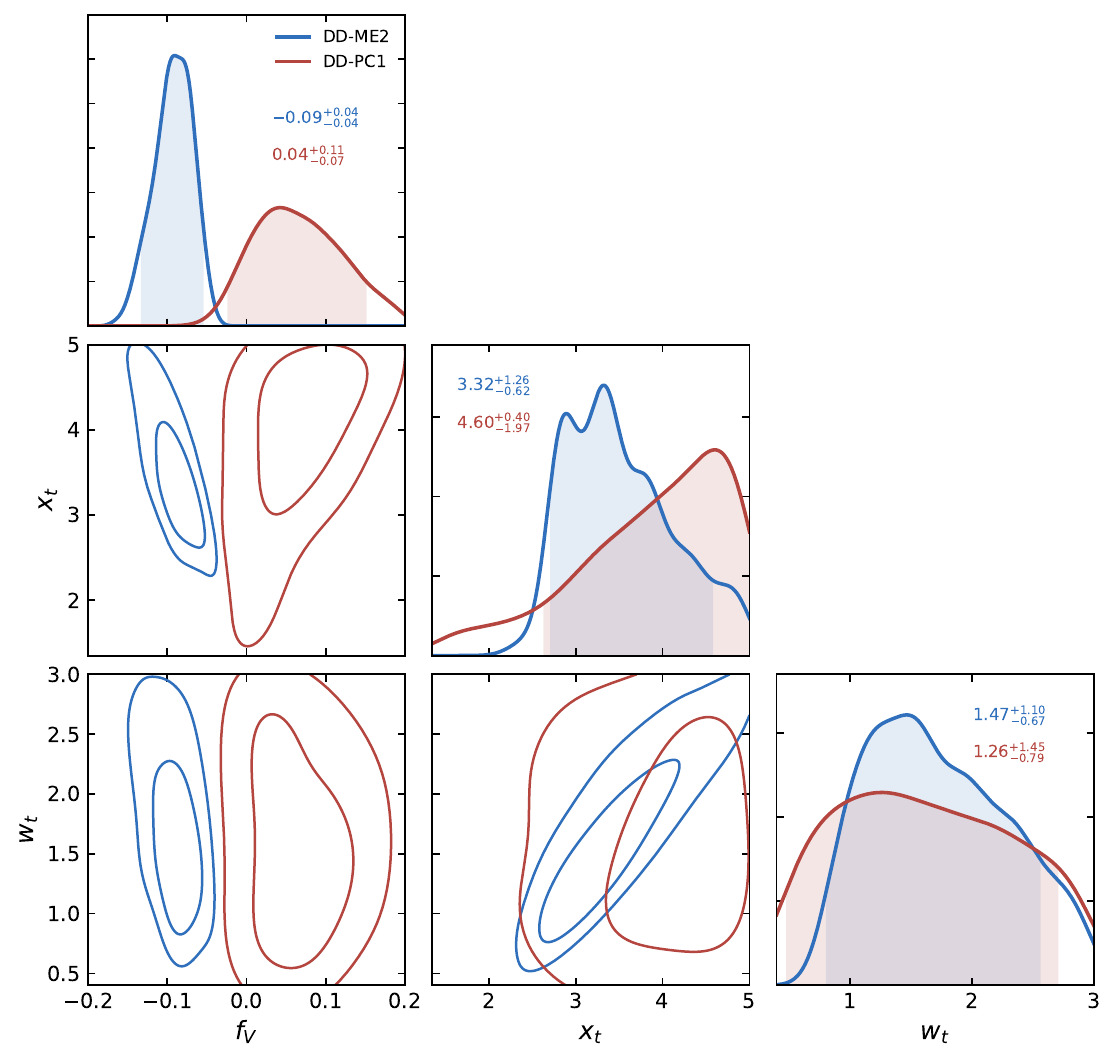}
  \caption{Posterior distribution of the high-density reshaping parameters
  \(\fv\), \(\xt\), and \(\wt\).  The diagonal panels show the one-dimensional
  marginalized posterior distributions together with the 90\%
  highest-posterior-density intervals; the plotted axis limits coincide with
  the uniform prior ranges.  Off-diagonal contours enclose the central 68\%
  and 90\% credible regions.  All three reshaping parameters are
  dimensionless.}
  \label{fig:corner}
\end{figure*}

Figure~\ref{fig:corner} shows the posterior distribution of the reshaping
parameters.  The \ddme posterior selects a well localized deformation strength
and a broad high-density onset window.  The one-dimensional posterior-mode
summaries are 
\begin{equation}
  \fv=-0.094^{+0.040}_{-0.040},\qquad
  \xt=3.31^{+1.27}_{-0.61},\qquad
  \wt=1.54^{+1.03}_{-0.74},
\end{equation}
where the uncertainties denote 90\% highest-posterior-density intervals.
The \ddpc posterior is much less
selective in the same parameter space, consistent with its role as a
reference functional that already lies closer to the preferred
multimessenger region.  This differs from generic EOS reconstructions, where
the posterior freedom is distributed over pressure, sound-speed, or spectral
parameters~\cite{Legred2021,Altiparmak2022,Brandes2023,Jiang2023}; here the
deformation is tied to one CDF channel and can be compared directly with the
original functional.

\begin{figure}[!t]
  \centering
  \includegraphics[width=0.86\linewidth]{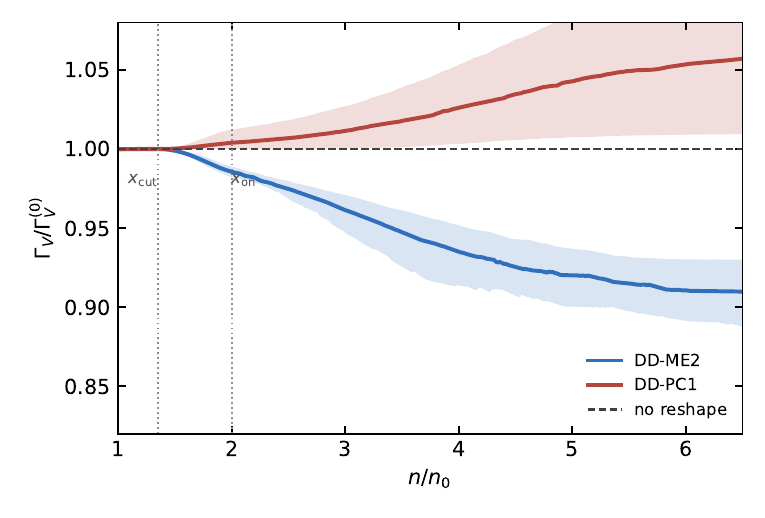}
  \caption{Posterior deformation of the isoscalar-vector coupling.  Solid
  curves show medians and shaded regions show central 68\% credible bands of
  \(\Gamma_V/\Gamma_V^{(0)}=1+\fv S(x;\xt,\wt)\) for the causal-filtered
  reshaped models.  The dashed horizontal line denotes the original
  no-reshape coupling, while the vertical dotted lines mark \(x_{\rm cut}\)
  and \(x_{\rm on}\).}
  \label{fig:vector-deformation}
\end{figure}

The corresponding posterior deformation of the vector coupling is shown in
Fig.~\ref{fig:vector-deformation}.  For \ddme, the causal-filtered posterior
selects a net reduction of the high-density vector coupling, with
\(\Gamma_V/\Gamma_V^{(0)}=0.91^{+0.02}_{-0.02}\) at \(6.5n_0\).  The median
therefore corresponds to an approximately \(8\)--\(10\%\) reduction of the
original \ddme vector coupling once the switch is active at several times
saturation density.  Thus the
large evidence gain is not a generic broadening of the EOS, but a diagnosis
that the original \ddme vector sector is too repulsive at high density for
the adopted likelihood.  The \ddpc posterior, in contrast, allows a mild
increase of the vector coupling but does not gain evidence after the prior
volume and causal filter are included.

\begin{figure}[!t]
  \centering
  \includegraphics[width=0.86\linewidth]{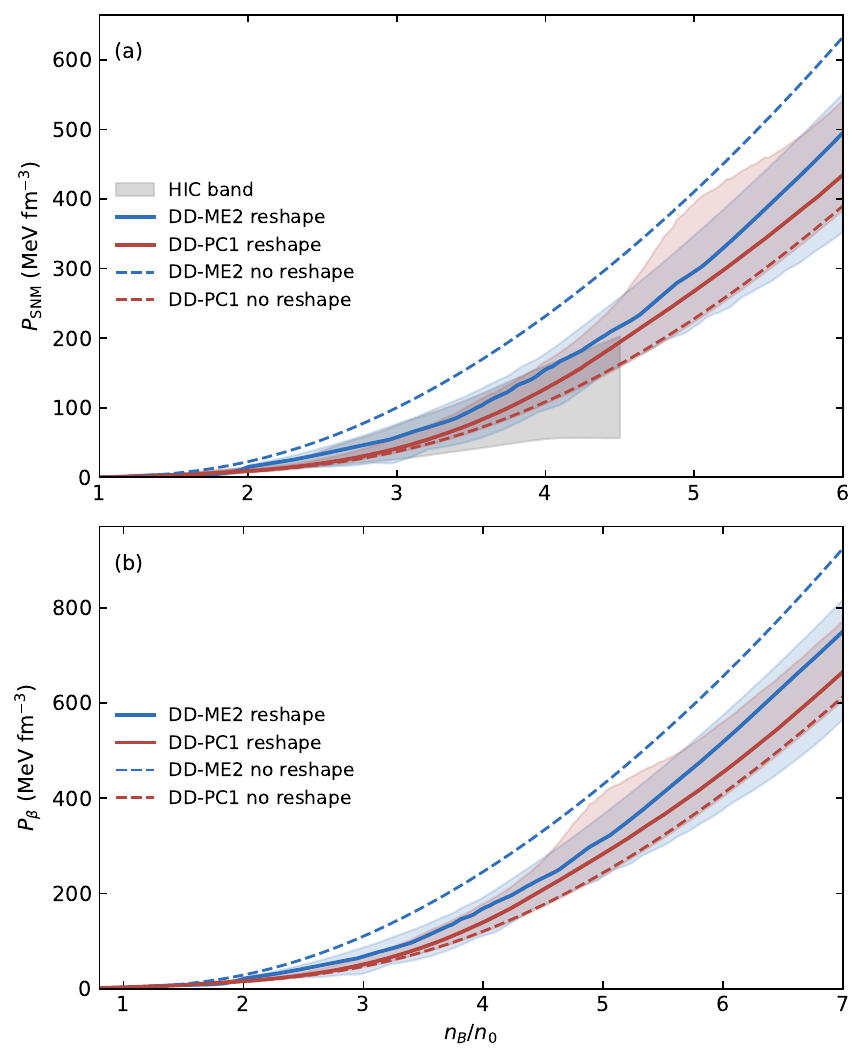}
  \caption{Posterior pressure bands.  Upper panel: symmetric nuclear matter
  compared with the HIC flow constraint of Ref.~\cite{Danielewicz2002}.
  Lower panel: beta-equilibrated matter obtained from the same Bayesian
  weights.  Dashed curves denote the original no-reshape functionals.}
  \label{fig:pressure}
\end{figure}

The impact of this deformation on the pressure is displayed in
Fig.~\ref{fig:pressure}.  In symmetric nuclear matter, the original \ddme EOS
is too stiff compared with
the HIC flow band, while the protected vector-channel extension shifts the
posterior pressure into the allowed region without changing finite-nucleus
observables.  The dashed no-reshape curves in Fig.~\ref{fig:pressure} make
this displacement explicit and show that the evidence gain is driven by a
real correction to the original high-density pressure rather than by a
cosmetic broadening of the posterior band.  This is the primary driver of the
large Bayes factor for the \ddme extension.  The same posterior weights also
determine the pressure of
beta-equilibrated matter, which connects the nuclear-matter pressure
constrained by HIC data to the stellar structure calculation.  The \ddpc
no-reshape curve and posterior band remain much closer to each other,
explaining why the additional parameter volume is not rewarded for \ddpc.
This comparison is consistent with recent joint HIC and astrophysical
analyses, which show that flow information is especially restrictive for the
pressure at a few times saturation density~\cite{Huth2022,Sorensen2024PPNP,
Tsang2024NatAstron}.

\begin{figure}[!t]
  \centering
  \includegraphics[width=0.86\linewidth]{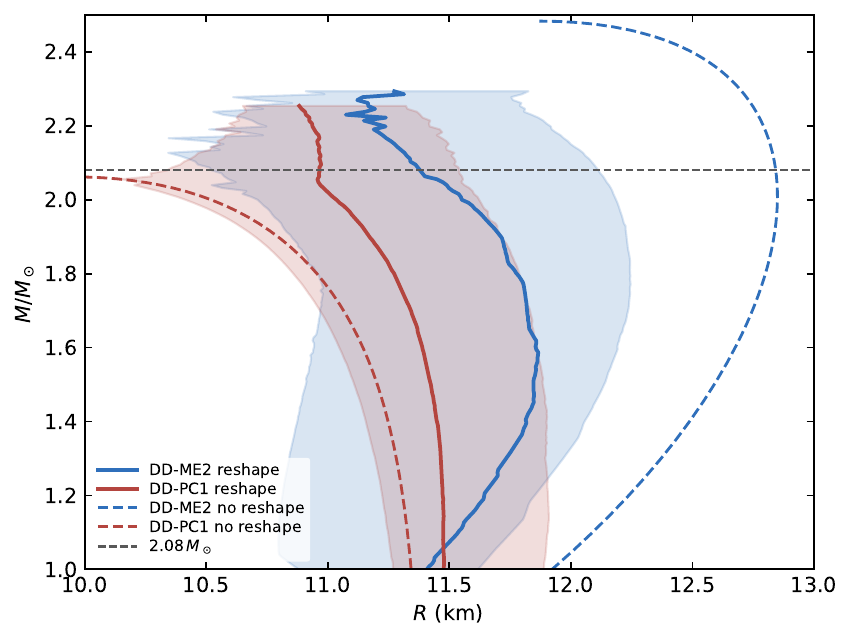}
  \caption{Posterior mass-radius relation obtained from the common
  HIC+\(M_{\max}\)+NICER+GW170817 likelihood.  Dashed curves denote the
  original no-reshape functionals.  The one-dimensional comparison with
  NICER-informed observables is summarized in Fig.~\ref{fig:observables}.}
  \label{fig:mr}
\end{figure}

\begin{figure}[!t]
  \centering
  \includegraphics[width=0.86\linewidth]{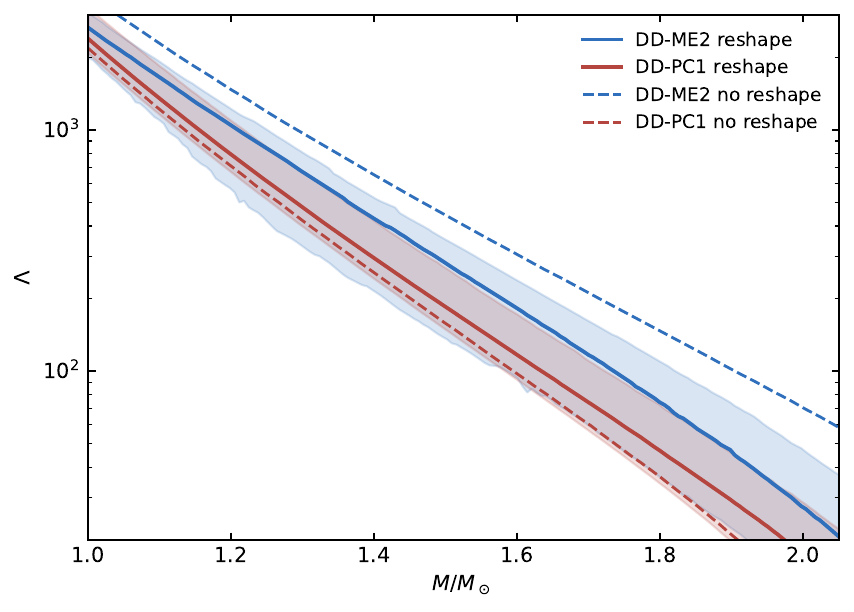}
  \caption{Posterior tidal deformability as a function of neutron-star mass.
  The tidal curves use the regularized Love-number solver for both families.
  Dashed curves denote the original no-reshape functionals.  The GW170817
  comparison is summarized in the \(\Lambda_{1.4}\) panel of
  Fig.~\ref{fig:observables}.}
  \label{fig:lambda}
\end{figure}

Figures~\ref{fig:mr} and~\ref{fig:lambda} show the corresponding mass-radius
and tidal-deformability predictions.  The same posterior weights are used
throughout, so the pressure, mass-radius, and tidal constraints are treated
consistently.  Compared with the dashed no-reshape baselines, the reshaped
\ddme posterior moves to smaller radii and lower tidal deformabilities while
remaining compatible with the maximum-mass constraint.  The \ddpc case shows
a more modest displacement, again consistent with its smaller Bayes-factor
gain.  The mass-radius posterior remains compatible with the
two-solar-mass requirement and with the NICER radius scale, while the
tidal-deformability posterior is shifted into the range preferred by
GW170817.  Thus the HIC-driven high-density modification does not solve one
constraint at the expense of the neutron-star observables; it produces a
common posterior compatible with all components of the likelihood.
The resulting radii lie in the same qualitative range favored by NICER
analyses of PSR J0740+6620 and PSR J0437--4715~\cite{Riley2021,Miller2021,
Choudhury2024,Rutherford2024}, while the tidal behavior remains compatible
with the GW170817 bounds~\cite{Abbott2018PRL}.  Compared with broad
multimessenger EOS reconstructions~\cite{Raaijmakers2021,Huth2022,Pang2024},
the present analysis reaches this compatibility through a
finite-nucleus-protected deformation of a specific CDF rather than through a generic EOS
parameterization.

\begin{figure*}[!t]
  \centering
  \includegraphics[width=0.92\linewidth]{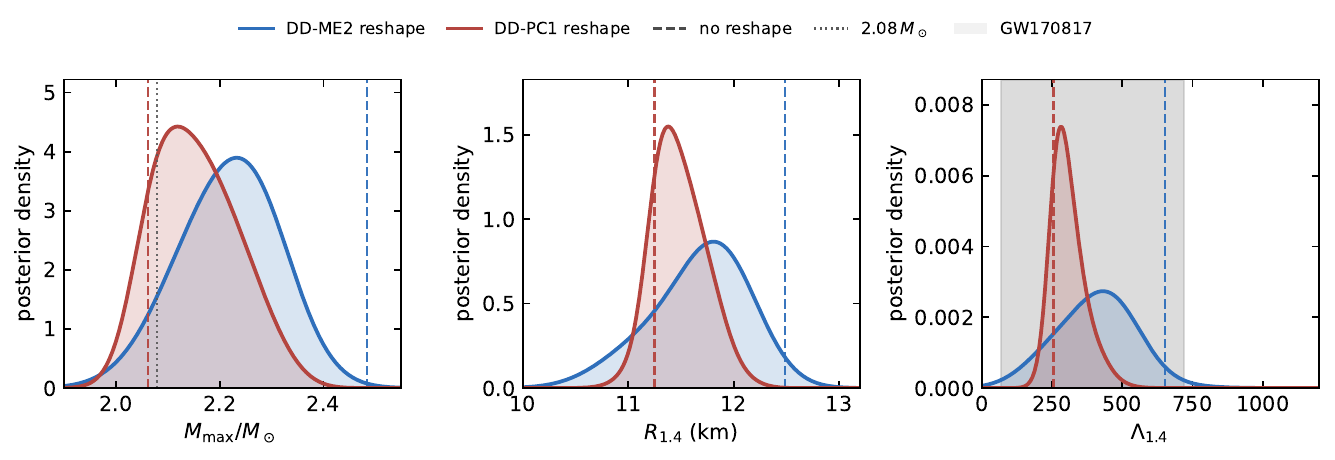}
  \caption{Posterior distributions of \(M_{\max}\), \(R_{1.4}\), and
  \(\Lambda_{1.4}\).  Solid curves and shaded regions show the reshaped
  posterior distributions, dashed vertical lines denote the original
  no-reshape baselines, the dotted line marks the \(2.08\,M_\odot\) maximum
  mass reference, and the gray band indicates the GW170817-informed
  \(\Lambda_{1.4}\) reference range from Ref.~\cite{Abbott2018PRL}.}
  \label{fig:observables}
\end{figure*}

The same comparison can be compressed into directly observable quantities, as
shown in Fig.~\ref{fig:observables}.  The dashed vertical lines mark the
original no-reshape functionals, while the colored distributions are the
reshaped posterior predictions.  For \ddme, the high-density extension moves
the excessively large no-reshape maximum mass, radius, and tidal
deformability toward the region selected by the combined HIC and neutron-star
likelihood.  For \ddpc, the shifts are smaller and the posterior remains
closer to the no-reshape baseline, consistent with the weaker evidence for
additional high-density flexibility.  This figure therefore makes explicit
that the evidence gain for \ddme is tied to simultaneous improvement in the
stellar observables, not only to the HIC pressure band.

\begin{figure}[!t]
  \centering
  \includegraphics[width=0.86\linewidth]{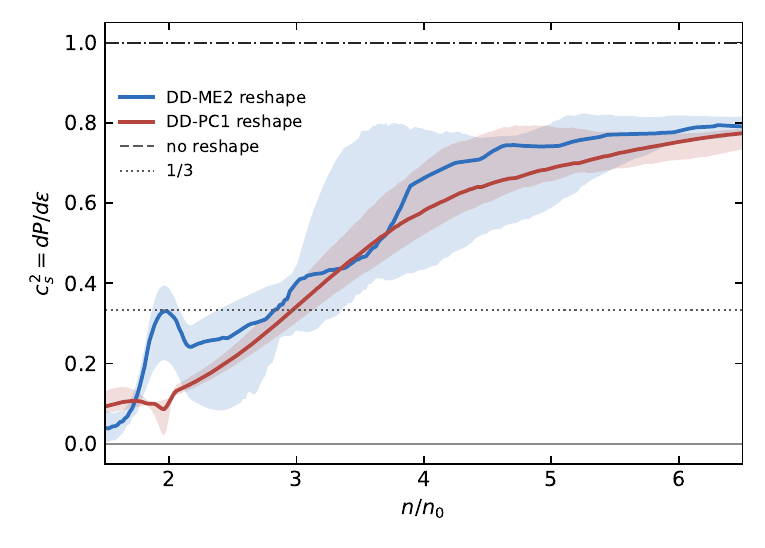}
  \caption{Posterior diagnostic for the squared sound speed in
  beta-equilibrated matter.  Solid curves show posterior medians and shaded
  regions show central 68\% credible bands for the causal-filtered reshaped
  models.  Dashed curves denote the two original no-reshape baselines; they
  are close on the scale of the plot.  The
  horizontal dotted and dash-dotted lines mark \(c_s^2=1/3\) and the causal
  limit \(c_s^2=1\), respectively.}
  \label{fig:cs2}
\end{figure}

As a microscopic consistency diagnostic, Fig.~\ref{fig:cs2} shows the
squared sound speed \(c_s^2=dP/d\varepsilon\) in beta-equilibrated matter.
The derivative is evaluated from the tabulated EOS by a local linear fit in
\(P(\varepsilon)\), which reduces sensitivity to point-to-point numerical
noise.  We impose the causal/stability filter
\(-0.05\le c_s^2\le1\) over \(1.5\le n/n_0\le6.5\), where the small negative
tolerance accounts for residual numerical differentiation oscillations.
Both no-reshape baselines satisfy this condition.  Without this filter, a
non-negligible part of the reshaped \ddpc posterior explores narrow
late-onset switches with \(c_s^2>1\).  The final evidence and posterior bands
therefore use the causal-filtered prior volume.  The resulting posterior
medians and central 68\% bands in Fig.~\ref{fig:cs2} remain below the causal
limit for both families, while the Bayes factor continues to select the
protected extension only for \ddme.  The preferred posteriors exceed the
conformal value \(c_s^2=1/3\) over part of the neutron-star density range.
We do not impose \(1/3\) as a bound, since the densities considered here are
not asymptotically perturbative; instead, the excess indicates the strong
interactions required in this hadronic CDF representation to satisfy the
combined HIC and neutron-star constraints.  This interpretation is consistent
with recent trace-anomaly analyses, in which a peak or excess in \(c_s^2\) at
intermediate densities is not by itself a violation of asymptotic
conformality, but can reflect a rapid density evolution of
\(\Delta=1/3-P/\varepsilon\)~\cite{Fujimoto2022PRL}.  Since collective-flow
extractions constrain an averaged stiffness of cold dense matter rather than
the identical beta-equilibrated EOS shown in Fig.~\ref{fig:cs2}, we use them
only as an external thermodynamic benchmark, not as an additional likelihood
term~\cite{Li2026TraceAnomalyFlow}.

The evidence comparison is summarized in Table~\ref{tab:evidence}.  The
table reports both the absolute evidence \(\ln Z\) entering the four-model
comparison and the within-family Bayes factor
\(\ln K_{\rm reshape}=\ln Z_{\rm reshape}-\ln Z_{\rm no\,reshape}\).
The original high-density sector of \ddme is incompatible with the adopted
multimessenger likelihood unless an additional finite-nucleus-protected
completion is introduced, giving
\(\ln K_{\rm reshape}=26.67\).  By contrast, the same extension gives
\(\ln K_{\rm reshape}=-0.44\) for \ddpc.  This contrast is central to the
interpretation: the Bayesian analysis does not simply reward extra
flexibility, but selects the high-density completion only when demanded by
the data.  In the four-model comparison the largest model weight is assigned
to the original \ddpc interaction, while the reshaped \ddme model carries the
only significant \ddme probability.  The result should therefore be read as a
controlled diagnosis of the functional dependence: the original \ddme vector
sector is too repulsive at high density for this likelihood, whereas \ddpc
does not require the same completion under the present data set.
This agrees with the broader lesson of recent Bayesian RMF/CDF studies:
current astrophysical data can strongly constrain high-density behavior, but
the conclusion remains model dependent unless the finite-nucleus calibration
and high-density extension are kept under explicit control~\cite{Salinas2023}.
The present two-family comparison should therefore be read as a controlled
\ddme--\ddpc demonstration rather than as a statistical sample of all CDFs;
extensions to DD-MEX, DD-LZ1, PC-PK1, and other modern families are left for
future work.

As a prior-volume check, Table~\ref{tab:prior-sensitivity} repeats the
evidence estimate on restricted subdomains of the same likelihood grid.  The
absolute grid evidence and Bayes factors in this table are deterministic
grid estimates, whereas Table~\ref{tab:evidence} reports the final MultiNest
evidence used for model comparison.  The grid estimate is used only as a fast
diagnostic, but the conclusion is stable:
the \ddme Bayes factor remains large and positive, whereas the \ddpc Bayes
factor remains negative for all tested prior windows.

\begin{table}[!t]
  \centering
  \caption{Grid-level prior sensitivity of the within-family Bayes factors.
  The baseline uses \(\fv\in[-0.20,0.20]\), \(\xt\in[1.35,5.00]\), and
  \(\wt\in[0.40,3.00]\).}
  \label{tab:prior-sensitivity}
  \begin{tabular}{lrr}
    \toprule
    prior window & \(\ln K_{\ddme}\) & \(\ln K_{\ddpc}\) \\
    \midrule
    baseline & 26.83 & -0.60 \\
    \(\xt\le4,\ \wt\le2\) & 27.46 & -1.01 \\
    \(\xt\le4\) & 27.04 & -1.17 \\
    \(\wt\le2\) & 27.07 & -0.46 \\
    \(|\fv|\le0.15\) & 27.09 & -0.43 \\
    \bottomrule
  \end{tabular}
\end{table}

We also tested whether the large \ddme Bayes factor is an artifact of the
chosen finite-nucleus cutoff scale.  Table~\ref{tab:xcut-sensitivity}
repeats the \ddme evidence estimate on a common sparse grid for
\(x_{\rm cut}=1.20,1.35,\) and \(1.50\), with \(x_{\rm on}=2.00\) fixed and
the same causal/stability filter imposed.  The evidence remains decisive in
all three cases, showing that the conclusion that the original \ddme
high-density sector strongly prefers a protected completion within the
adopted likelihood is not tied to the specific value \(x_{\rm cut}=1.35\).

\begin{table}[!t]
  \centering
  \caption{Sensitivity of the \ddme grid evidence to the finite-nucleus
  cutoff \(x_{\rm cut}\).  The estimates use a common sparse grid in
  \((\fv,\xt,\wt)\), fixed \(x_{\rm on}=2.00\), the
  HIC+\(M_{\max}\)+NICER+GW170817 likelihood, and the causal/stability
  filter \(-0.05\le c_s^2\le1\) over \(1.5\le n/n_0\le6.5\).}
  \label{tab:xcut-sensitivity}
  \begin{tabular}{lrrrr}
    \toprule
    \(x_{\rm cut}\) & \(x_{\rm on}\) & causal fraction & \(\ln Z\) & \(\ln K_{\ddme}\) \\
    \midrule
    1.20 & 2.00 & 0.541 & -17.342 & 28.21 \\
    1.35 & 2.00 & 0.463 & -17.795 & 27.76 \\
    1.50 & 2.00 & 0.375 & -19.116 & 26.44 \\
    \bottomrule
  \end{tabular}
\end{table}

The evidence result is also sensitive, as expected, to how strongly the HIC
flow band is imposed.  To quantify this dependence without changing any EOS
calculation, Table~\ref{tab:hic-sensitivity} recomputes the causal grid
evidence after inflating the HIC one-sided Gaussian width by factors of 1.5
and 2.0, while keeping the massive-pulsar, NICER, GW170817, and causal
filters unchanged.  The same table also changes the HIC density interval
while leaving the remaining likelihood terms fixed.  The \ddme Bayes factor
decreases from 27.58 to 11.88 and 6.54 when the HIC width is inflated by
factors of 1.5 and 2.0, respectively, but remains positive and sizable.  It
also remains positive for the lower-density, higher-density, and
intermediate-density HIC subwindows, with \(\ln K_{\ddme}=17.65\), 21.21,
and 9.73.  Removing the HIC term almost eliminates the preference.  Thus the
large \ddme evidence gain is driven by a real tension between the original
high-density pressure and the HIC stiffness constraint, but it is not an
artifact of the nominal digitized HIC-band width or of using the full
\(1.3\le n/n_0\le4.5\) HIC interval.  The \ddpc reference case remains close
to neutral under the same tests.

\begin{table}[!t]
  \centering
  \caption{Grid-level sensitivity to the HIC likelihood prescription.  The
  upper block changes the width of the one-sided Gaussian HIC penalty over
  the default \(1.3\le n/n_0\le4.5\) interval.  The lower block keeps the
  nominal HIC width but changes the density interval.  All other likelihood
  terms and the causal/stability filter are kept fixed.  The ``no HIC'' row
  removes the HIC contribution from the same grid likelihood and is shown
  only as a diagnostic of which data component drives the evidence gain.}
  \label{tab:hic-sensitivity}
  \begin{tabular}{lrr}
    \toprule
    HIC treatment & \(\ln K_{\ddme}\) & \(\ln K_{\ddpc}\) \\
    \midrule
    default interval, nominal width & 27.58 & -0.07 \\
    default interval, width \(\times 1.5\) & 11.88 & -0.01 \\
    default interval, width \(\times 2.0\) & 6.54 & 0.03 \\
    no HIC term & 0.34 & 0.09 \\
    \midrule
    \(1.3\le n/n_0\le3.0\) & 17.65 & 0.04 \\
    \(2.0\le n/n_0\le4.5\) & 21.21 & -0.03 \\
    \(3.0\le n/n_0\le4.5\) & 9.73 & -0.02 \\
    \bottomrule
  \end{tabular}
\end{table}

Finally, Table~\ref{tab:future-mass} illustrates a prospective high-mass
scenario motivated by GW190814.  The secondary component of GW190814 has a
mass of \(2.50\)--\(2.67\,M_\odot\)~\cite{Abbott2020GW190814}, but its
nature remains unsettled: it may be a low-mass black hole, a rapidly rotating
neutron star, or, under more speculative assumptions, an extremely massive
nonrotating neutron star.  Several studies have explored the neutron-star
interpretation and its implications for dense-matter EOSs
~\cite{Most2020GW190814,ZhangLi2020GW190814,Fattoyev2020GW190814}, while
recent multimessenger inference gives
\(M_{\rm TOV}=2.25^{+0.08}_{-0.07}\,M_\odot\)~\cite{Fan2024MTOV}.  We
therefore do not include GW190814 in the baseline likelihood.  Instead, the
\(2.08\,M_\odot\) maximum-mass term is replaced by a one-sided lower-bound
constraint at a hypothetical confirmed nonrotating neutron-star mass
\(M_{\rm obs}\), using \(\sigma_M=0.07\,M_\odot\).  This is not part of the
baseline evidence in Table~\ref{tab:evidence}.  Rather, it shows how the same
diagnostic framework would respond if a substantially heavier nonrotating
neutron star were established.  The \ddme extension remains strongly
preferred throughout, but the \ddpc Bayes factor turns positive and grows
with \(M_{\rm obs}\), because the data would then favor a stiffening
high-density deformation of \ddpc.

\begin{table}[!t]
  \centering
  \caption{Prospective high-mass scenario using grid evidence.  The table
  reports within-family Bayes factors after replacing the current
  \(2.08\,M_\odot\) maximum-mass term by a one-sided lower-bound likelihood
  centered at \(M_{\rm obs}\), with \(\sigma_M=0.07\,M_\odot\).  The
  \(2.59\,M_\odot\) row is motivated by the secondary mass scale of
  GW190814 and should be read as a scenario test, not as a baseline
  assumption.}
  \label{tab:future-mass}
  \begin{tabular}{lrr}
    \toprule
    \(M_{\rm obs}\) & \(\ln K_{\ddme}\) & \(\ln K_{\ddpc}\) \\
    \midrule
    2.08 & 27.58 & -0.07 \\
    2.20 & 27.45 & 1.34 \\
    2.30 & 26.98 & 4.09 \\
    2.40 & 25.55 & 7.83 \\
    2.50 & 22.81 & 12.14 \\
    2.59 & 20.30 & 16.33 \\
    \bottomrule
  \end{tabular}
\end{table}

\begin{table*}[!t]
  \centering
  \caption{Bayesian evidence for the four-model comparison using the common
  HIC+\(M_{\max}\)+NICER+GW170817 likelihood and the causal/stability filter
  \(-0.05\le c_s^2\le1\) over \(1.5\le n/n_0\le6.5\) for the reshaped
  models.  The column \(\ln K_{\rm reshape}\) gives the Bayes factor of each
  reshaped model relative to its own no-reshape baseline; it is therefore
  reported only for the reshape rows.  The run-to-run scatter
  \(\sigma_{\rm run}\) and mean internal nested-sampling error
  \(\langle\sigma_{\ln Z}\rangle\) summarize the three independent MultiNest
  runs for the reshaped models; they are not applicable to the
  zero-parameter baselines.}
  \label{tab:evidence}
  \begin{tabular}{lrrrrrrr}
    \toprule
    model & \(k\) & \(\ln Z\) & \(\sigma_{\rm run}\) &
    \(\langle\sigma_{\ln Z}\rangle\) & \(\Delta\ln Z\) &
    \(\ln K_{\rm reshape}\) & weight \\
    \midrule
    \ddme & 0 & -45.555 & -- & -- & -31.157 & -- & \(1.78\times10^{-14}\) \\
    \ddme + reshape & 3 & -18.887 & 0.195 & 0.039 & -4.489 & 26.67 & 0.0068 \\
    \ddpc & 0 & -14.399 & -- & -- & 0.000 & -- & 0.6049 \\
    \ddpc + reshape & 3 & -14.842 & 0.289 & 0.012 & -0.443 & -0.44 & 0.3883 \\
    \bottomrule
  \end{tabular}
\end{table*}

\section{Summary and outlook}

We have introduced a finite-nucleus-protected high-density extension of
covariant density functionals and applied it to the isoscalar-vector channel
of \ddme and \ddpc.  The extension leaves finite-nucleus observables
unchanged within numerical precision and is tested against HIC flow, massive
pulsars, NICER mass-radius measurements, and GW170817.

The Bayesian evidence diagnoses a strong tension in the original \ddme
high-density sector and indicates that an additional protected completion is
strongly preferred within the adopted likelihood, with \(\ln K=26.67\)
relative to the original interaction after the causal/stability filter is
imposed.  The \ddpc reference case shows the opposite behavior: the same
extension is not favored once the
parameter-volume penalty and causal prior volume are included.  The proposed
finite-nucleus-protected construction therefore provides a controlled way to
separate robust finite-nucleus calibration from high-density EOS inference,
and it identifies when a given covariant density functional requires an
additional high-density vector-channel completion.  The present work should
therefore be viewed as a controlled diagnostic step toward future global CDF
analyses, rather than as a completed unified refit.  Finite-nucleus data are
protected rather than refitted here because the purpose is to isolate the
high-density sector while preserving the original finite-nucleus
calibration; they will be incorporated directly into the likelihood in a
future global analysis.

Several extensions follow naturally from this diagnostic framework.  First,
finite-nucleus masses, charge radii, and neutron-skin observables should be
included explicitly in a combined likelihood so that the protection strategy
can be replaced by a fully simultaneous calibration.  Second, the same
Bayesian model-selection test should be applied to additional CDF families
and to a larger model space in which the high-density isovector-vector
channel and alternative switching functions are allowed to vary.  Third, the
thermodynamic consequences of the protected completion should be audited with
dimensionless diagnostics such as \(c_s^2\) and
\(\Delta=1/3-P/\varepsilon\), treating beta-equilibrated neutron-star matter
and HIC-relevant nuclear matter separately before making direct comparisons
with recent astrophysical and collective-flow extractions.  These steps will
clarify which parts of the present evidence result are specific to \ddme and
\ddpc, and which reflect a more general tension between finite-nucleus CDF
calibrations and high-density multimessenger constraints.

\section*{Data availability}

The EOS tables, likelihood grids, MultiNest posterior samples, and plotting
scripts used to produce the figures and evidence tables are available from
the authors upon reasonable request and will be deposited in a public
repository upon publication.

\FloatBarrier

\end{document}